\documentstyle[aps,prd,preprint,tighten,amssymb,eqsecnum,newlfont]{revtex}
\newcommand{\be}{\begin{equation}}
\newcommand{\ee}{\end{equation}}
\newcommand{\sslash}{s\hspace{-2.0mm} /}
\newcommand{\kslash}{k\hspace{-2.1mm} /}
\begin{document}
\preprint{
\begin{tabular}{r}
UWThPh-2000-21\\
May 2000
\end{tabular}
}
\draft 
\title{Elastic neutrino -- electron scattering of solar neutrinos and
potential effects of magnetic and electric dipole moments}
\author
{W. Grimus\footnote{grimus@doppler.thp.univie.ac.at} and
T. Schwetz\footnote{schwetz@doppler.thp.univie.ac.at}}
\address
{Institute for Theoretical Physics, University of Vienna,\\
Boltzmanngasse 5, A--1090 Vienna, Austria} 

\maketitle
\begin{abstract}
We consider elastic neutrino -- electron scattering of solar neutrinos
with magnetic moments and electric dipole moments, where the solar
neutrino state at the scattering site is determined by the evolution
in matter and solar magnetic fields of the initial electron neutrino
state. We present the general cross section for an arbitrary
superposition of active and sterile neutrino types with positive and
negative helicities, with particular emphasis on the effect of transverse
polarization, which gives rise to an azimuthal asymmetry as a function of the
recoil electron momentum. Within our physically motivated
approximation, we perform a general CP analysis and show that in the
1-Dirac and 2-Majorana neutrino cases no CP-violating effects are
present, which means that it is not possible to distinguish between
magnetic and electric dipole moments in these cases. We also study the 
consequences of neutrino energy averaging on the cross section and
stress that in the 2-Majorana neutrino case this averaging leads to a
suppression of the transverse neutrino polarization effects.
\end{abstract}

\pacs{PACS numbers: 14.60.St, 13.40.Em, 14.60.Pq}

\section{Introduction}

Recently, the physics of neutrino oscillations \cite{osc}
has become one of the most active fields of research in particle
physics. One of the reasons is the experimental evidence for
neutrino oscillations found in atmospheric neutrino measurements \cite{sk}.
The other important problem in this field is the longstanding solar
neutrino deficit which also finds 
a natural explanation in terms of neutrino oscillations, whether by
vacuum oscillations or by the MSW effect \cite{matter} (for recent works see
\cite{suzuki,fits}). Concerning the solar neutrino puzzle, it has been
noticed long time ago that neutrino magnetic moments (MM) and/or
electric dipole moments (EDM) of order $10^{-11}\, \mu_B$, where $\mu_B$ is
the Bohr magneton, and a sizable magnetic field in the solar interior
can contribute to a solution of this problem \cite{cisneros,MM,okun}
(for reviews see also Ref.\cite{MMreview}). 

A particular attractive scenario in this context, which combines the
matter effect with the effect of solar magnetic fields and neutrino
MMs or EDMs, is given by resonant
spin -- flavour precession (RSFP) \cite{LM,Akhmedov,minakata}, which allows
for good fits of the solar neutrino data 
(for recent papers see \cite{ALP,GN,semi-torr,PA,DT,BPRS,miranda}). 
This scenario is
possible even without neutrino mixing. However, very little is known
about magnetic fields in the solar interior and one has to
resort to plausible assumptions, a problem in all attempts to
solve the solar neutrino puzzle with neutrino MMs and EDMs.

If neutrinos have MMs and/or EDMs then the electromagnetic interaction
will give a contribution to elastic neutrino  -- electron scattering in
addition to the weak interactions \cite{em}, which is enhanced
at low energies. This observation has
been used to derive laboratory limits on neutrino MM/EDMs
\cite{caso}. Independently, such limits can also be obtained by taking
advantage of the solar neutrino flux \cite{mourao,beacom}. Furthermore,
if MM/EDMs of solar neutrinos and solar magnetic fields are important
for a solution of the solar neutrino puzzle, then the solar neutrino
flux on earth will have some transverse polarization in general.
It has been stressed that such a polarization leads to a
weak-electromagnetic interference cross section in elastic neutrino --
electron scattering, which could be observed via an azimuthal
asymmetry in the distribution of the electron recoil momenta
in the plane orthogonal to the incoming neutrino momentum 
\cite{BF,VE,GS94,Semikoz,PSSV98,PSSV99}. A suitable experiment to find
such an effect could be the HELLAZ experiment \cite{HELLAZ}.

It is the purpose of this paper to present a general and consistent
study of elastic neutrino -- electron scattering of solar neutrinos,
thereby incorporating the twofold effects of the neutrino MM/EDMs:
1.\ They contribute to the evolution of the initial electron neutrino
state with negative helicity to the state which is detected on earth
by elastic neutrino -- electron scattering.
2.\ Neutrino MM/EDMs contribute to elastic neutrino -- electron
scattering by providing a pure electromagnetic
cross section and a weak-electromagnetic interference cross section.
We will assume an arbitrary number of neutrinos, including 
neutrinos of the sterile type, and derive the cross section for an arbitrary
superposition of neutrino flavours or types and helicities. We will
meticulously distinguish between Dirac and Majorana neutrinos.
We will employ the following approximation in our physical scenario: we
neglect all neutrino masses in the elastic neutrino -- electron
cross section, but retain the usual term quadratic in the neutrino
masses in the neutrino evolution equation \cite{matter}. This has to
be kept in mind when assessing the results of this paper.

Since we neglect neutrino masses in scattering, the most adequate basis for
our consideration is the flavour basis of neutrino states, though
physically, as we will show, no basis is preferred. We,
therefore, put particular emphasis on the formulation of the MM and EDM
matrices in the flavour basis and stress that the most useful entity in this
context is the MM/EDM matrix $\lambda = \mu - id$, where $\mu$ and $d$
are the MM and EDM matrices, respectively. Considering $\lambda$ and
the neutrino mixing matrix $U_L$, we make a general discussion
of the independent, physical phases in our problem. We will show that
there are no such phases in the 1-Dirac and 2-Majorana neutrino cases,
from which it follows that in these cases no distinction between MM and EDM
within our approximation is possible. Finally, we discuss decoherence
effects as a consequence of neutrino energy averaging, which is of importance
for the weak-electromagnetic interference cross section and neutrino
mass-squared differences larger than around $10^{-10}$ eV$^2$. This
effect is caused by vacuum oscillations between sun and earth and by
the inevitable energy averaging due to finite energy and angle
resolution in the detection of the recoil electron in $\nu\, e^-$ scattering.

The paper is organized as follows. In Section \ref{MMEDM} we discuss
the MM and EDM interaction for Dirac and Majorana neutrinos. The
evolution equation of the solar neutrino state in matter and magnetic
fields is explained in Section \ref{evoleq}. Section \ref{density}
treats the formulation of the neutrino density matrix, which is
applied in Section \ref{elastic} in the calculation of the elastic
neutrino -- electron cross section. Section \ref{solar nus} contains
the applications of Sections \ref{evoleq} and \ref{elastic} for the
1-Dirac and 2-Majorana cases and a general discussion of the physical
phases and the effect of CP invariance in our problem.
Section \ref{decoherence} discusses the decoherence effect due to
neutrino energy averaging and in Section \ref{conclusions} we present
our conclusions.

\section{Neutrino magnetic moments and electric dipole moments}
\label{MMEDM}

If Dirac neutrinos are furnished with magnetic moments (MM) and electric
dipole moments (EDM), the interaction with the electromagnetic field is
described by the Hamiltonian
\be\label{HD}
\mathcal{H}_{\mathrm{em}}^D = 
\frac{1}{2} \bar{\nu} (\mu + i d \gamma_5)
\sigma^{\alpha \beta} \nu F_{\alpha \beta} =
\frac{1}{2} \bar{\nu}_R \lambda
\sigma^{\alpha \beta} \nu_L F_{\alpha \beta} + \mathrm{h.c.} \,.
\ee
We assume for the time being that we consider neutrinos flavours (or types if
take into account sterile fields $\nu_s$). Thus,
$\nu^T = (\nu_e , \nu_\mu , \nu_\tau , \nu_s , \ldots )$ is the vector
of the flavour eigenfields. Hermiticity of the Hamiltonian (\ref{HD}) requires
that the MM and EDM matrices are both hermitian:
\be\label{dagger}
\mu^\dagger = \mu, \quad d^\dagger = d \,.
\ee
Therefore, the diagonal elements of $\mu$ and $d$ are real for Dirac
neutrinos. With the decomposition $\nu = \nu_L + \nu_R$, where $\nu_L$ and
$\nu_R$ are left and right-chiral fields, respectively, the second part of
Eq.(\ref{HD}) follows, where the MM and EDM matrices are condensed in
the non-hermitian matrix
\be\label{lambda}
\lambda = \mu - i d 
\quad \mbox{with} \quad 
\mu = \frac{1}{2} ( \lambda + \lambda^\dagger ) \,, \;
d   = \frac{i}{2} ( \lambda - \lambda^\dagger ) \,.
\ee

The Hamiltonian (\ref{HD}) can be rewritten in any basis. One might, e.g.,
want to formulate in the basis of neutrino mass eigenfields. 
In the most general case one has to perform separate unitary rotations
\be\label{unitary} 
\nu_{L} = S_L \nu'_L \quad \mbox{and} \quad \nu_{R} = S_R \nu'_R
\ee
on the left and right-chiral
fields, respectively, in which case the matrix (\ref{lambda}) transforms as 
\be\label{basis}
\lambda' = S_R^\dagger \lambda S_L \,.
\ee
According to Eq.(\ref{lambda}), the MM and EDM matrices in the new
basis are obtained by
\be\label{newD}
\renewcommand{\arraystretch}{1.5}
\begin{array}{rcl}
\mu' & = & \frac{1}{2} \left\{ S_R^\dagger \mu S_L + S_L^\dagger \mu S_R -
i ( S_R^\dagger d S_L - S_L^\dagger d S_R ) \right\}, \\
d' & = & \frac{1}{2} \left\{ S_R^\dagger d S_L + S_L^\dagger d S_R +
i ( S_R^\dagger \mu S_L - S_L^\dagger \mu S_R ) \right\} \,,
\end{array}
\ee
respectively. Note that the matrix $\lambda$ is the object which transforms
simply under a basis change, not the MM and EDM matrices.

For Majorana neutrinos we have the Hamiltonian \cite{SV}
\be\label{HM}
\mathcal{H}_\mathrm{em}^M = - \frac{1}{4} \nu^T C^{-1} (\mu + i d \gamma_5)
\sigma^{\alpha \beta} \nu F_{\alpha \beta} =
- \frac{1}{4} \nu_L^T C^{-1} \lambda
\sigma^{\alpha \beta} \nu_L F_{\alpha \beta} + \mathrm{h.c.} \,,
\ee
where $C$ is the charge conjugation matrix and $\nu = \nu_L + (\nu_L)^c$. The
superscript $c$ denotes the charge conjugate field. 
By the anticommutation properties of the
fermionic fields $\nu_L$, it follows that
\be
\mu^T = -\mu, \quad d^T = -d \,.
\ee
Thus the MM and EDM matrices are antisymmetric and hermitian, and,
therefore, imaginary. Then, $\lambda$ (\ref{lambda}), which is defined 
as in the Dirac case, is antisymmetric as well. 
The factor $-1/4$ in the Hamiltonian (\ref{HM})
has been chosen because it would appear by rewriting the Dirac form
(\ref{HD}) in purely left-handed fields, i.e., in Majorana from.
Since the right-handed component of a Majorana field is the
charge conjugate of the left-handed component, left and right basis rotations
are related by 
\be
S_R = S_L^* \,,
\ee
and in the new basis we have
\be\label{newM}
\mu' = i \left\{ 
\mathrm{Im} (S_L^T \mu S_L ) - \mathrm{Re} (S_L^T d S_L ) \right\}, \;
d' = i \left\{
\mathrm{Im} (S_L^T d S_L ) + \mathrm{Re} (S_L^T \mu S_L ) \right\} \,.
\ee

In which basis the MM and EDM matrices are the ``physically relevant'' ones
depends on the situation. Obviously, if one could experimentally distinguish
neutrino mass eigenstates, $\mu$ and $d$ in the mass eigenbasis would be
considered physical and diagonal MMs and EDMs could be distinguished in the
Dirac case. For transition moments the freedom of making phase rotations,
i.e., basis transformations 
with $S_L = S_R$ being a diagonal matrix of phase factors, blurs the
distinction between transition MMs and EDMs even for Dirac neutrino mass
eigenstates. For Majorana neutrinos in the mass eigenbasis, however, 
there is only the freedom of making a transformation with $S_L$ being a
diagonal sign matrix. In practice one cannot measure neutrino mass eigenstates
and we will consider the situation that in elastic neutrino -- electron
scattering all neutrino masses are neglected, but neutrino masses enter in the
usual quadratic way in the evolution equation of the neutrino state with
background matter and magnetic fields. In Section \ref{solar nus} we will 
come back to the question of physically observable quantities related to MMs
and EDMs in this context.

\section{The evolution equation of the solar neutrino state}
\label{evoleq}

If neutrinos possess a sizeable MM and/or EDM it is possible that solar
neutrinos acquire a transverse polarization on their way to the surface
of the sun if a large solar magnetic field exists
\cite{BF,VE,PSSV98,PSSV99,Semikoz}. The evolution of the neutrino
state produced in the core of the sun
under the influence of the solar magnetic field and matter effects
is governed by the Schr\"odinger-like equation 
\cite{matter,SV,LM,Akhmedov,minakata,GScharn,balaji}
\be\label{evol}
i\frac{d}{dz} \left(\begin{array}{c} \varphi_- \\ \varphi_+
\end{array}\right) = \left(\begin{array}{cc}
V_L + \frac{1}{2\omega}M^\dagger M & -B_+ \lambda^\dagger \\
-B_-\lambda & V_R + \frac{1}{2\omega}M M^\dagger
\end{array}\right)
\left(\begin{array}{c} \varphi_- \\ \varphi_+  \end{array}\right) 
\equiv H_\mathrm{eff}
\left(\begin{array}{c} \varphi_- \\ \varphi_+  \end{array}\right) \,.
\ee
In this equation, $\varphi_-$  and $\varphi_+$
denote the vectors of neutrino flavour
wave functions corresponding to negative and positive helicity,
respectively, and $\omega$ denotes the neutrino energy. The elements of
$\varphi_\mp$ are ordered according to $\alpha = e, \mu, \tau$, followed by
an arbitrary number of sterile neutrinos, in general.
The matter potential $V_L$ \cite{matter} is given by
\be
V_L = \sqrt{2}\, G_F\, \mbox{diag}(n_e - n_n/2, -n_n/2, -n_n/2, 0, \ldots)
\,,
\ee
where $n_e\:(n_n)$ is the electron (neutron) density in the
sun. $M$ denotes the neutrino mass matrix in the flavour basis. With
the diagonal matrix $\hat m$ of neutrino masses and the unitary
diagonalizing matrices $U_L$ and $U_R$, we have the relations
\be\label{masses}
\hat{m} = U_R^\dagger M U_L \Rightarrow 
M^\dagger M = U_L \hat{m}^2 U_L^\dagger 
\quad \mbox{and} \quad
M M^\dagger = U_R \hat{m}^2 U_R^\dagger \,.
\ee
Furthermore, we use the definition
\be
B_\pm = B_x \pm iB_y \,.
\ee
Note that in Eq.(\ref{evol}) the neutrino propagates along the $z$-axis
and in our approximation only $B_x$ and $B_y$ -- the components
of the solar magnetic field orthogonal to the neutrino momentum -- 
contribute to the neutrino evolution.

Let us list the important differences between Dirac and Majorana neutrinos 
with respect to the quantities appearing in the
evolution equation (\ref{evol}):
\be\label{diff}
\begin{array}{rcccc}
\setlength{\arraycolsep}{8mm}
\mbox{Dirac neutrinos:} & V_R = 0, & M\: \mbox{arb.}, & U_R\: \mbox{arb.}, & 
\lambda\: \mbox{arb.}, \\
\mbox{Majorana neutrinos:} & V_R = -V_L, & M^T = M, & U_R = U_L^*, &
\lambda^T = -\lambda. 
\end{array}
\ee
In this table arb. stands for arbitrary.

The neutrinos are produced as electron neutrinos in the sun at the
coordinate $z_0$ and are detected on earth at $z_1$. Hence we express
the initial condition as
\be\label{initial}
\varphi_-(z_0) = \left(
\begin{array}{c} 1 \\ 0 \\ \vdots \end{array} \right), \,
\varphi_+(z_0) = \left(
\begin{array}{c} 0 \\ 0 \\ \vdots \end{array} \right) \,,
\ee
and the neutrino state at the detector is formally given by
\be\label{final}
\left( \begin{array}{c} a_- \\ a_+ \end{array} \right) \equiv
\left( \begin{array}{c} \varphi_-(z_1) \\ \varphi_+(z_1) \end{array}
\right) =
P \exp \left\{ -i \int_{z_0}^{z_1} \!\! dz\, H_\mathrm{eff}(z)
\right\} 
\left( \begin{array}{c} 1 \\ 0 \\ 0 \\ \vdots \end{array} \right) \,.
\ee
In this equation $P$ denotes path ordering.
For a given magnetic field along the neutrino path in the sun,
the neutrino state described by the vectors $a_\mp$ can in principle
be obtained by solving Eq.(\ref{evol}), as function of neutrino MMs,
EDMs, masses and mixing parameters. The densities $n_e$ and $n_n$
are provided by the Solar Standard Model.
The flavour vectors $a_\mp$ have to be used in the calculation of
the elastic neutrino -- electron cross section of solar neutrinos.

\section{The density matrix and polarization vectors}
\label{density}

In the calculation of the neutrino scattering 
cross section we need the density matrix of the initial neutrino state
\be
\rho^{\alpha\beta} = \sum_{r,s = \pm} u_r(k)\bar{u}_s(k)
\rho^{\alpha\beta}_{rs} \,.
\ee
The elements of the density matrix obey the relations
$\sum_{\alpha, r} \rho^{\alpha\alpha}_{rr} = 1$ and
$\rho^{\alpha\beta}_{rs} = ( \rho^{\beta\alpha}_{sr} )^*$.
In the case of full coherence they are connected with the coefficients
$a_\mp^\alpha$ (\ref{final}) by
\be\label{rhoa}
\rho^{\alpha\beta}_{rs} = a^\alpha_r {a^\beta_s}^* \,.
\ee
In the Dirac representation of the gamma matrices we have for the 4-spinor
$u$ for a massless neutrino
\be\label{u}
u_\pm(k) = \sqrt{\omega} \left(\begin{array}{c} \chi_\pm \\ \pm\chi_\pm
\end{array}\right)\,,
\ee
where $\omega$ is the energy of the initial neutrino.
Now we choose the $z$-axis along the direction of the initial neutrino
momentum. Hence the neutrino 4-momentum is 
$k=(\omega, 0, 0, \omega)^T$ and the 2-spinors of positive and
negative helicity are given by
\be\label{chi}
\chi_+ = \left( \begin{array}{c} 1 \\ 0 \end{array} \right)\,,\quad
\chi_- = \left( \begin{array}{c} 0 \\ 1 \end{array} \right)\,,
\ee
respectively.
Using Eqs.(\ref{u}) and (\ref{chi}) it is easy to verify that the density
matrix can be written in the following two equivalent ways:
\begin{eqnarray}
\rho^{\alpha\beta} &=& \frac{1}{2} \left\{
(1+\gamma_5)\rho^{\alpha\beta}_{++} + (1-\gamma_5)\rho^{\alpha\beta}_{--} +
\frac{1+\gamma_5}{2} \sslash^{\alpha\beta}_{+-} +
\frac{1-\gamma_5}{2} \sslash^{\alpha\beta}_{-+} \right\} \kslash
\label{rho1}\\
&=& \frac{1}{2} \left\{
\rho^{\alpha\beta}_{++} + \rho^{\alpha\beta}_{--} +
\gamma_5 (\rho^{\alpha\beta}_{++} - \rho^{\alpha\beta}_{--}) +
\gamma_5 \sslash^{\alpha\beta}_\bot \right\} \kslash \,, \label{rho2}
\end{eqnarray}
with the generalized polarization 4-vectors orthogonal to the neutrino
momentum
\be\label{s}
s^{\alpha\beta}_{+-} = \left( \begin{array}{c} 0 \\ \rho^{\alpha\beta}_{+-} \\
i\rho^{\alpha\beta}_{+-} \\ 0 \end{array}\right) \,,\quad
s^{\alpha\beta}_{-+} = \left( \begin{array}{c} 0 \\ -\rho^{\alpha\beta}_{-+} \\
i\rho^{\alpha\beta}_{-+} \\ 0 \end{array}\right) \,,\quad
s^{\alpha\beta}_\bot = s^{\alpha\beta}_{+-} - s^{\alpha\beta}_{-+} =
\left( \begin{array}{c} 0 \\
\rho^{\alpha\beta}_{+-} + \rho^{\alpha\beta}_{-+} \\
i(\rho^{\alpha\beta}_{+-} - \rho^{\alpha\beta}_{-+}) \\ 0
\end{array}\right) \,.
\ee
The longitudinal component of the polarization can be read off from 
Eq.(\ref{rho2}) and is given by
$\rho^{\alpha\beta}_{++} - \rho^{\alpha\beta}_{--}$. The polarization
vectors of a flavour mixed neutrino state are complex in general with
$( s^{\alpha\beta}_{-+} )^* = -s^{\beta\alpha}_{+-}$
and $( s^{\alpha\beta}_\bot )^\ast = s^{\beta\alpha}_\bot$.
Only in the case $\alpha = \beta$, the vector $s^{\alpha\alpha}_\bot$ is
real. In Ref.\cite{PSSV99} an expression similar to (\ref{rho1}) is
used for the density matrix. In the case of a single neutrino flavour
Eq.(\ref{rho2}) reduces to the expressions given in Refs.\cite{GS94,bilenky}.

\section{Elastic -- neutrino electron scattering with arbitrary 
neutrino polarization} 
\label{elastic}

In this section we present the weak, electromagnetic and interference
cross sections for the process 
\be
\nu(k) + e^-(p) \to \nu(k') + e^-(p') \,.
\ee
We work in the restframe of the initial electron and use the following
notation for the 4-momenta: 
\be
k = \left( \begin{array}{c} \omega \\ \vec{k} \end{array} \right) \,,
\; p = \left( \begin{array}{c} m_e \\ \vec{0} \end{array} \right) \,,\;
k' = \left( \begin{array}{c} \omega - T \\ {\vec{k}}^\prime \end{array}
\right) \,,\; 
p' = \left( \begin{array}{c} m_e + T \\ \vec{p}\,' \end{array} \right)
\,,
\ee 
where $m_e$ is the electron mass, neutrino masses are 
neglected in the cross sections
($k^2 = {k'}^2 = 0$) and $T = E_e'-m_e$ is the recoil energy of the
scattered electron.
For the angle $\theta = \angle (\vec{p}\,', \vec{k})$ of the recoil
electron, momentum conservation implies 
\be
\cos\theta = \frac{\omega + m_e}{\omega} \sqrt{\frac{T}{T+2m_e}} \,,
\ee
and the electron recoil energy $T$ is bounded by
$0 \le T \le T_{\mathrm{max}}$ with
$T_{\mathrm{max}} = 2\omega^2 / (2\omega + m_e)$.

The cross section for elastic neutrino -- electron scattering consists
of three terms
\be\label{cross}
\frac{d^2 \sigma}{dT d \phi} = \frac{d^2 \sigma_\mathrm{w}}{dT d \phi} +
\frac{d^2 \sigma_\mathrm{em}}{dT d \phi} + 
\frac{d^2 \sigma_\mathrm{int}}{dT d \phi} \,,
\ee
where $\phi$ is the azimuthal angle which is measured in the plane
orthogonal to the momentum of the initial neutrino.
The first and the second term are the pure weak and electromagnetic terms,
respectively, and the third term is the interference term between the weak
and the electromagnetic amplitude.

\subsection{The weak cross section}

The weak interaction of neutrinos with electrons is described by
the effective Hamiltonian
\be\label{HW}
\mathcal{H}_\mathrm{w} = \frac{G_F}{\sqrt{2}} \sum_\alpha
\bar{\nu}_\alpha \gamma_\lambda (1 - \gamma_5)\nu_\alpha \:
\bar{e} \gamma^\lambda (g_V^\alpha - g_A^\alpha \gamma_5) e \,,
\ee
where $G_F$ is the Fermi constant and
\be\label{gvga}
\begin{array}{cclccl}
g_V^e & = & 2 \sin^2\Theta_W + 1/2, & g_A^e & = & 1/2, \\
g_V^{\mu,\tau} & = & 2 \sin^2\Theta_W - 1/2, & 
g_A^{\mu,\tau} & = & -1/2, \\
g_V^s & = & 0, & g_A^s & = & 0, 
\end{array}
\ee
with the weak mixing angle $\Theta_W$. To write down the weak
cross section it is useful to define the left and right-handed constants
\be\label{grgl}
g_L^\alpha = \frac{1}{2} \left( g_V^\alpha + g_A^\alpha \right) \,,\quad
g_R^\alpha = \frac{1}{2} \left( g_V^\alpha - g_A^\alpha \right) \,,
\ee
respectively. Note that for active neutrinos, independent of the flavour
$\alpha$,  we have $g_R^\alpha = \sin^2\Theta_W$,
but, of course, $g_R^\alpha=0$ for sterile neutrinos.

With these constants,
the weak cross sections for Dirac and Majorana neutrinos are given by
\begin{eqnarray}
\frac{d^2 \sigma^D_\mathrm{w}}{dT d \phi} &=& \sum_\alpha
|a_-^\alpha|^2 \frac{d^2 \sigma(\nu_\alpha e^-)}{dT d \phi} \,, 
\label{csWD} \\
\frac{d^2 \sigma^M_\mathrm{w}}{dT d \phi} &=& \sum_\alpha \left(
|a_-^\alpha|^2 \frac{d^2 \sigma(\nu_\alpha e^-)}{dT d \phi} +
|a_+^\alpha|^2
\frac{d^2 \sigma(\bar{\nu}_\alpha e^-)}{dT d \phi} \right) \,,
\label{csWM}
\end{eqnarray}
respectively, where
\begin{eqnarray}
\frac{d^2 \sigma(\nu_\alpha e^-)}{dT d \phi} &=&
\frac{G_F^2 m_e}{\pi^2} \left( (g_L^\alpha)^2 +
(g_R^\alpha)^2 \left(1-\frac{T}{\omega} \right)^2 -
g_L^\alpha g_R^\alpha \frac{m_e T}{\omega^2} \right) \,, 
\label{csWnu} \\
\frac{d^2 \sigma(\bar{\nu}_\alpha e^-)}{dT d \phi} &=&
\frac{G_F^2 m_e}{\pi^2} \left( (g_R^\alpha)^2 +
(g_L^\alpha)^2 \left(1-\frac{T}{\omega} \right)^2 -
g_L^\alpha g_R^\alpha \frac{m_e T}{\omega^2} \right)
\label{csWantinu}
\end{eqnarray}
are the cross sections for elastic scattering of neutrinos and antineutrinos
of flavour $\alpha$ off electrons, respectively.
$|a_-^\alpha|^2$ ($|a_+^\alpha|^2$)
is the probability of finding a left-handed (right-handed) neutrino of flavour
$\alpha$ in the initial state. The right-handed states correspond to 
antineutrinos in the Majorana case, 
whereas for Dirac neutrinos they do not interact
weakly and hence the respective term is absent in Eq.(\ref{csWD}).

\subsection{The electromagnetic cross section}

The electromagnetic cross section has the same form for
Dirac and Majorana neutrinos:
\be\label{csEM}
\frac{d^2 \sigma_\mathrm{em}}{dT d\phi} = \frac{\alpha^2}{2m_e^2 \mu_B^2}
\left( \frac{1}{T} - \frac{1}{\omega} \right)
\left( a_-^\dagger \lambda^\dagger \lambda a_- +
a_+^\dagger \lambda \lambda^\dagger a_+ \right) \,,
\ee
where $\alpha = e^2 / 4\pi$ and $\mu_B = e/ 2m_e$. The matrix $\lambda$
is given in Eq.(\ref{lambda}) and we have
\be\label{ll}
\lambda^\dagger \lambda = \mu^2 + d^2 - i [\mu, d] \,,\quad
\lambda \lambda^\dagger = \mu^2 + d^2 + i [\mu, d]\,.
\ee
Eq.(\ref{csEM}) reduces to the well known result in the single flavour
Dirac case \cite{em}, for special cases with several flavours see
Refs.\cite{GS98,beacom}.  

\subsection{The interference cross section}

If the polarization of the initial neutrino possesses a component
transverse to its momentum, an interference term between the
weak and electromagnetic amplitude appears in the cross section \cite{BF,VE}.
This term shows a dependence on the azimuthal angle $\phi$. With the
definitions
\be\label{defg}
g^\alpha = g_V^\alpha \left(2 - \frac{T}{\omega} \right) +
g_A^\alpha \frac{T}{\omega} \:,\quad
\bar{g}^\alpha = g_V^\alpha \left(2 - \frac{T}{\omega} \right) -
g_A^\alpha \frac{T}{\omega}
\ee
and $\hat{k} = \vec{k} / |\vec{k}|$, we find for Dirac and Majorana
neutrinos, respectively,
\begin{eqnarray}
\frac{d^2 \sigma_\mathrm{int}^D}{dT d\phi} &=&
\frac{G_F \alpha}{2 \sqrt{2}\pi m_e T \mu_B} \sum_{\alpha,\beta} \, 
\mbox{Re} \left[
\left( \vec{p}\,' \mu_{\alpha\beta}+ (\hat{k} \times \vec{p}\,')
d_{\alpha\beta} \right) g^\alpha \vec{s}\,_{+-}^{\beta\alpha} \right] \,,
\label{csintD1}\\
\frac{d^2 \sigma_\mathrm{int}^M}{dT d\phi} &=&
\frac{G_F \alpha}{2 \sqrt{2}\pi m_e T \mu_B} \sum_{\alpha,\beta}
\mbox{Re} \left[
\left( \vec{p}\,' \mu_{\alpha\beta}+ (\hat{k} \times \vec{p}\,')
d_{\alpha\beta} \right) \left( g^\alpha \vec{s}\,_{+-}^{\beta\alpha} -
\bar{g}^\alpha \vec{s}\,_{-+}^{\beta\alpha} \right)
\right] \,. \label{csintM1} 
\end{eqnarray}
With $\vec{p}\,' = (p'_x, p'_y, p'_z)^T$ and $\hat{k} = (0,0,1)^T$, one has
$\hat{k} \times \vec{p}\,' = (-p'_y , p'_x , 0)^T$
and, using the explicit form of the polarization vectors (\ref{s}) 
given by
\be
\vec{s}\,^{\beta\alpha}_{+-} =
\left( \begin{array}{c} a_+^\beta a_-^{\alpha\ast} \\
ia_+^\beta a_-^{\alpha\ast} \\ 0 \end{array}\right) \:,\quad
\vec{s}\,^{\beta\alpha}_{-+} =
\left( \begin{array}{c} -a_-^\beta a_+^{\alpha\ast} \\
ia_-^\beta a_+^{\alpha\ast} \\ 0 \end{array}\right) 
\ee
for full coherence, we obtain
\begin{eqnarray}
\frac{d^2 \sigma_\mathrm{int}^D}{dT d\phi} &=&
F\, \mbox{Re}\left[a_+^\dagger \lambda g a_- (p'_x - ip'_y)\right] \,,
\label{csintD2}\\
\frac{d^2 \sigma_\mathrm{int}^M}{dT d\phi} &=&
F\, \mbox{Re}\left[a_+^\dagger (\lambda g + \bar{g}\lambda)a_-
(p'_x - ip'_y)\right] \,, \label{csintM2}
\end{eqnarray}
where we have defined $F = G_F \alpha /2 \sqrt{2}\pi m_e T \mu_B$ and
the diagonal matrices
$g = \mbox{diag}(g^\alpha )$ and $\bar{g} = \mbox{diag}(\bar g^\alpha )$.

For a single flavour Dirac neutrino, Eq.(\ref{csintD1}) reduces to the
expression given in Ref.\cite{GS94}.

\section{Elastic neutrino -- electron scattering of solar neutrinos}
\label{solar nus}

\subsection{A single Dirac neutrino}

For a single Dirac neutrino the Hamiltonian governing the evolution
equation (\ref{evol}) has the form
\begin{equation}\label{H1D}
H_{\mathrm{eff}} = \left(
\begin{array}{cc} V_L + \frac{m^2}{2\omega} & -B_+ (\mu + id) \\
-B_- (\mu - id) & \frac{m^2}{2\omega}
\end{array} \right) = 
\mathrm{diag} ( 1, e^{-i\delta} )\, H'_\mathrm{eff} \,
\mathrm{diag} ( 1, e^{i\delta} )
\end{equation} 
with 
\begin{equation}
\mu + i d = \sqrt{\mu^2 + d^2}\, e^{i\delta} \,,
\end{equation}
where $H'_\mathrm{eff}$ differs from $H_\mathrm{eff}$ by
$\sqrt{\mu^2 + d^2}$ instead of $\mu \pm id$. 
This leads to
\be
a_+ = e^{-i\delta} a'_+ \,,
\ee
where $a'_+$ and also $a_-$ do not depend on $\delta$ but only 
on $\sqrt{\mu^2 + d^2}$.

In this case we have
\be\label{aa}
a_+^\dagger \lambda g a_- = \sqrt{\mu^2 + d^2}\, (a'_+)^* a_- g^e
\quad \mbox{and} \quad
a_-^\dagger \lambda^\dagger \lambda a_- + 
a_+^\dagger \lambda \lambda^\dagger a_+ = \mu^2 + d^2 \,.
\ee
Therefore, one cannot distinguish between the MM and EDM of a single Dirac
neutrino \cite{okun}
in elastic neutrino -- electron scattering of solar neutrinos.

As a function of the azimuthal angle, the interference cross section
(\ref{csintD2}) has a maximum when $p'_x + i p'_y$ and 
$a_+^\dagger \lambda g a_-$ are aligned in the complex plane. In general, with
matter effects in addition to the magnetic field interaction, there is no
obvious meaning of the phase of the latter quantity. However, if we can
neglect matter effects and the direction of the magnetic field is fixed,
i.e., 
\be\label{B}
B_+(z) = B(z) e^{i\beta} \,,
\ee
where $\beta$ does not depend on $z$, the solution of the evolution equation
(\ref{evol}) for a single Dirac neutrino is given by \cite{MM}
\be
a_- = \cos \left( \sqrt{\mu^2 + d^2} \int_{z_0}^{z_1} dz B(z) \right)
\quad \mbox{and} \quad
a_+ = i e^{-i(\delta + \beta)} \sin \left( \sqrt{\mu^2 + d^2}
\int_{z_0}^{z_1} dz B(z) \right) \,.
\ee
We have left out the irrelevant phase stemming from $m^2/2\omega$.
Consequently, we obtain 
\be\label{int1D}
2\, a_+^\dagger \lambda g a_- = -ie^{i\beta} \sqrt{\mu^2 + d^2}\, g^e
\sin \left( 2\sqrt{\mu^2 + d^2} \int_{z_0}^{z_1} dz B(z) \right) \,,
\ee
and the maximum of the interference cross section at
\be
\left( \begin{array}{c} p'_x \\ p'_y \end{array} \right) \propto
\left( \begin{array}{r} B_y \\ -B_x \end{array} \right)
\ee
defines an azimuthal angle orthogonal to the direction of the transverse
magnetic field. Thus, in such a situation the interference cross
section allows to determine the direction of the magnetic field.

\subsection{Two Majorana neutrinos}
\label{2M}

Now we come to the second simplest case, which is the case of two Majorana
neutrinos with flavours $e$ and $x$. For two Majorana flavours
the matrix $\lambda$ is simply given by
\be\label{2Mlambda}
\lambda = \Lambda \epsilon = | \Lambda | e^{-i\delta} \epsilon
\quad \mbox{with} \quad
\epsilon = \left(\begin{array}{rr} 0 & 1 \\ -1 & 0 \end{array}\right)\,.
\ee
Discussing first the electromagnetic cross section (\ref{csEM}),
Eq.(\ref{2Mlambda}) readily gives \cite{beacom}
\be\label{em}
a_-^\dagger \lambda^\dagger \lambda a_- + 
a_+^\dagger \lambda \lambda^\dagger a_+ =
| \Lambda |^2 \,.
\ee
Due to conservation of probability and the simple form of the matrix
$\lambda$ (\ref{2Mlambda}), the coefficients $a_\mp$ do not occur
and the electromagnetic cross section depends thus only on
the single electromagnetic moment in the problem.
The term (\ref{em}) has the same structure as in the 1-Dirac case 
(see Eq.(\ref{aa})).

Now we will perform a thorough discussion of all possible phases
appearing in $U_L$ and $\lambda$ and we will show that they play no role 
in the two flavour case, in the framework of our approximation. 
The matrix $U_L$ (\ref{masses}) can be written as
\be\label{UL}
U_L = e^{i\hat{\sigma}} V e^{i\hat{\rho}} \,,
\ee
where $\hat{\sigma}$ and $\hat{\rho}$ are diagonal phase matrices and $V$
corresponds to the KM matrix, which is real for two flavours \cite{KM}.
Note that the phases $\rho_j$ play no role in our problem for any number of
flavours because these so-called Majorana phases \cite{Mphases}
drop out in the effective Hamiltonian of the evolution equation 
(\ref{evol}). Furthermore, a phase transformation with $e^{i\hat \sigma}$
upon $\lambda$ has the effect
\be\label{Lambda}
e^{i\hat{\sigma}} \lambda e^{i\hat{\sigma}} = \tilde{\Lambda} \epsilon
\quad \mbox{with} \quad 
\tilde{\Lambda} = \Lambda e^{i(\sigma_e + \sigma_x)} =
| \Lambda | e^{2i\gamma} \quad \mbox{and} \quad
\gamma = (-\delta + \sigma_e + \sigma_x)/2 \,.
\ee
These observations suggest to perform the phase transformation
\be\label{HP}
\mathcal{P}^\dagger H_\mathrm{eff} \mathcal{P} = H'_\mathrm{eff}
\quad \mbox{with} \quad
\mathcal{P} = \mathrm{diag}\, 
\left( e^{i\hat{\sigma}} e^{-i\gamma}, 
e^{-i\hat{\sigma}} e^{i\gamma} \right) \,,
\ee
where 
\be\label{Heff'}
H'_\mathrm{eff} = \left(
\begin{array}{cc} 
V_L + \frac{1}{2\omega} V \hat{m}^2 V^T & B_+ |\Lambda | \epsilon \\
-B_-|\Lambda | \epsilon & -V_L + \frac{1}{2\omega} V \hat{m}^2 V^T
\end{array} \right)
\ee
depends only on the orthogonal matrix $V$ and $|\Lambda |$. 
From the relation (\ref{HP}) it follows that
\be\label{a'}
a_- = e^{i\hat{\sigma}} e^{-i\gamma} a'_- \,, \quad
a_+ = e^{-i\hat{\sigma}}e^{i\gamma} a'_+ \,,
\ee
where $a'_\mp$ is obtained by evolution with $H'_\mathrm{eff}$ and is
independent of any of the phases discussed above.

With Eqs.(\ref{2Mlambda}) and (\ref{a'}) we arrive at
\be\label{int}
a_+^\dagger (\lambda g + \bar g \lambda) a_- = | \Lambda | \left\{
- (a_{\hphantom{\prime\hskip+0.5pt}+}^{\prime\, x})^* 
   a_{\hphantom{\prime\hskip+0.5pt}-}^{\prime\, e} 
  (g^e + \bar{g}^x)
+ (a_{\hphantom{\prime\hskip+0.5pt}+}^{\prime\, e})^* 
   a_{\hphantom{\prime\hskip+0.5pt}-}^{\prime\, x} 
  (\bar{g}^e + g^x) \right\} \,.
\ee
Thus, no phase of $U_L$ and $\lambda$ remains in the 
interference and electromagnetic cross sections of elastic neutrino --
electron scattering of solar neutrinos. In particular, no
distinction is possible between the transition MM and EDM, irrespective of the
basis where they are defined\footnote{This statement disagrees with the
result of Ref.\cite{PSSV99}.} 
(see also Eqs.(\ref{V}) and (\ref{lcs})). 
This statement is valid in the framework of the
approximation of neglecting neutrino masses in the scattering process and the
validity of the evolution equation (\ref{evol}).

Let us briefly mention three special cases within the 2-Majorana
neutrino case. If we have no neutrino mixing, only the first term in
Eq.(\ref{int}) contributes, because only $a^e_-$ and $a^x_+$ 
are non-zero. This is the minimal scenario for RSFP. 
Neutrino mixing also drops out of $H'_\mathrm{eff}$ (\ref{Heff'}), 
if we require $m_1 = m_2$. This case can be conceived as stemming from a
Zeldovich -- Konopinski -- Mahmoud neutrino $\nu = \nu_{eL}+(\nu_{xL})^c$
\cite{ZKM}, which has a conserved lepton number.  
If in addition to $m_1 = m_2$ we set $V_L = 0$ and
require a fixed direction of the transverse
magnetic field (see Eq.(\ref{B})), then we have an interference cross
section analogous to the 1-Dirac neutrino case, where in the expression
(\ref{int1D}) the quantity $g^e$ is replaced by $g^e + \bar g^x$ and
$\sqrt{\mu^2 + d^2}$ by $|\Lambda |$. Again, with the help of the
interference cross section, one could in principle determine the
direction of the magnetic field.

\subsection{Phase counting}
\label{phase counting}

Having discussed at length the 1-Dirac and 2-Majorana neutrino cases,
where we have shown that there are no physical phases, we now proceed
to the general phase counting for $n$ neutrinos. We will assume that
the charged lepton mass matrix is diagonal and positive. Our focus is
on the neutrino flavour fields or states in the left-handed sector
because we have formulated elastic neutrino -- electron scattering
with these entities.

In the case of Dirac neutrinos we have no interaction of the
right-handed neutrinos. Therefore, in the physical situation under
consideration, the matrix $U_R$ (\ref{masses}) is unphysical and can be
rotated away and the mass term has the form
\be\label{HDmass}
\mathcal{H}^D_\mathrm{mass} = \bar \nu_R M \nu_L + \mbox{h.c.}
\quad \mbox{with} \quad M = \hat{m} U_L^\dagger \,.
\ee
We are left with the following phase freedom:
\begin{eqnarray}
&& \nu_L = e^{i\sigma_L} \nu'_L, \quad \nu_R = e^{i\sigma_R} \nu'_R \quad
\Rightarrow \nonumber \\
&& \lambda \to e^{-i\sigma_R} \lambda\, e^{i\sigma_L}, \quad
M \to e^{-i\sigma_R} M e^{i\sigma_L} \quad \mbox{or} \quad
U_L \to e^{-i\sigma_L} U_L e^{i\sigma_R} \,.
\label{phasetrafoD}
\end{eqnarray}
The vectors $\varphi_\mp$ (\ref{evol}) and, therefore, also $a_\mp$, 
transform in the same way as the fields $\nu_{L,R}$. 
Obviously, the cross sections (\ref{csEM}) and (\ref{csintD2}) are invariant
under the phase transformation (\ref{phasetrafoD}). 
In general, $U_L$ is of the form (\ref{UL}),
where $V$ is of the KM type with $(n-1)(n-2)/2$ phases \cite{KM}. The phases
$\hat \rho$ drop out of $M^\dagger M$ and $M M^\dagger$, and with 
$\sigma_L = \hat \sigma$ we shift the phases $\hat \sigma$ to
$\lambda$ (see Eq.(\ref{phasetrafoD})). The 
remaining phase freedom in $\sigma_R$ is used to remove $n$ phases from
the $n^2$ phases in $\lambda$. Thus, we are left with 
$(n-1)(n-2)/2 + n(n-1) = (3n-2)(n-1)/2$ independent phases in the problem.

In the Majorana case\footnote{In order to have a MM/EDM for Majorana
neutrinos, we need $n \geq 2$.} we have the mass term
\be\label{HMmass}
\mathcal{H}^M_\mathrm{mass} = 
-\frac{1}{2} \nu_L^T C^{-1} M \nu_L + \mbox{h.c.}
\quad \mbox{with} \quad M = U_L^* \hat{m} U_L^\dagger
\ee
and the phase transformation
\begin{eqnarray}
&& \nu_L = e^{i\sigma_L} \nu'_L \quad
\Rightarrow \nonumber \\
&& \lambda \to e^{i\sigma_L} \lambda\, e^{i\sigma_L}, \quad
M \to e^{i\sigma_L} M e^{i\sigma_L} \quad \mbox{or} \quad
U_L \to e^{-i\sigma_L} U_L \,.
\label{phasetrafoM}
\end{eqnarray}
Again the electromagnetic cross section (\ref{csEM}) and the interference
cross section (\ref{csintM2}) are invariant under
this phase transformation, the phases $\hat \rho$ -- the Majorana phases --
cancel, and the phases $\hat \sigma$ can be transferred by the
transformation (\ref{phasetrafoM}) to $\lambda$, which is antisymmetric
and, therefore, has $n(n-1)/2$ phases. Now there is no freedom to
remove phases from the MM/EDM matrix as in the Dirac case, but one
phase of $\lambda$ can still be eliminated by by redefining $\nu_L$
with a common phase of the type of $\gamma$ in Eq.(\ref{HP}).
Finally, we arrive at 
$(n-1)(n-2)/2 + n(n-1)/2 -1 = n(n-2)$ independent phases.

In summary, we have found the following numbers of independent,
physical phases in the problem we are studying:
\be\label{number}
\left.
\begin{array}{rc}
\mbox{Dirac case:}    & (3n-2)(n-1)/2 \\
\mbox{Majorana case:} & n(n-2)
\end{array}
\right\} \; \mbox{physical phases.}
\ee
Eq.(\ref{number}) explains why in the 1-Dirac and 2-Majorana neutrino
cases we have no phases left in the elastic neutrino -- electron cross
section. It is interesting to observe that phases can be shifted
from the mixing matrix $U_L$ to the MM/EDM matrix $\lambda$ and vice
versa. In our physical setting the phases in the MM/EDM matrix are not
physical in the sense of the phase transformations
(\ref{phasetrafoD}) and (\ref{phasetrafoM}) and the transformed MMs
and EDMs (\ref{newD}) and (\ref{newM}) and, therefore, a distinction
between MM and EDM is thus unphysical as well.

\subsection{Invariance of the cross section under general basis
transformations} 
\label{invariance}

Up to now we have considered only the freedom of performing 
phase transformations on the neutrino fields.
However, since we neglect neutrino masses in the cross section, we
could use any basis for its calculation. Let us first consider Dirac
neutrinos and the general basis transformation (\ref{unitary}). Then
the transformed MM/EDM matrix is given by Eq.(\ref{basis}) and the
transformed flavour coefficients are obtained by
\be
a_- = S_L a'_- \quad \mbox{and} \quad a_+ = S_R a'_+ \,. 
\ee
We also have to take into account that in the weak Hamiltonian
(\ref{HW}) the transformed matrices
\be 
g_{V,A}' = S_L^\dagger g_{V,A} S_L
\ee 
appear, where the matrices 
$g_{V,A}$ are diagonal matrices of the coupling constants (\ref{gvga}).
Taking this set of transformed quantities, we can immediately rewrite the
cross sections (\ref{csEM}) and (\ref{csintD2}) in terms of the primed
quantities. The same can be done with the weak cross section
(\ref{csWD}), if we notice that it is a function of the 4 expressions
$a_-^\dagger g_{V,A} g_{V,A} a_-$ (see Eq.(\ref{csWnu})). 

For Majorana neutrinos, we have $S_R = S_L^*$. In addition, one can
easily check that in the calculation of the antineutrino part
(\ref{csWantinu}) of the weak cross section (\ref{csWM}) one gets
${g_{V,A}'}^*$ and the same applies to the $\bar g$ term in the
interference cross section (\ref{csintM2}). These observations lead to
invariance of the Majorana neutrino cross section under the general
basis transformation (\ref{unitary}). Consequently, in our physical
problem there is no preferred basis, neither for Dirac nor for
Majorana neutrinos. 

\subsection{CP invariance}

Let us now study the effect of CP invariance on the phases discussed
above. We focus on the MM and EDM matrices in the flavour basis 
and then compare with the mass basis. 
It will turn out that all the phases counted in the previous
subsection are effects of CP violation. 

In the flavour basis, CP
invariance for Dirac neutrinos is expressed as invariance of the
Lagrangian under the CP transformation
\begin{equation}\label{CP}
\begin{array}{clcl}
\mbox{CP:} & \nu_L & \to & -e^{i\alpha_L} C \nu_L^* \\
           & \nu_R & \to & -e^{i\alpha_R} C \nu_R^* \\
           & \ell  & \to & -e^{i\alpha_L} C \ell^*  \\
& F_{\alpha \beta} & \to & 
- \varepsilon(\alpha) \varepsilon(\beta) F_{\alpha \beta} \,,
\end{array}
\end{equation}
where $\varepsilon(\alpha)$ is 1 for $\alpha = 0$ and $-1$ for $\alpha
= 1,2,3$, $\ell$ is the vector of the charged lepton fields and $\alpha_L$,
$\alpha_R$ denote diagonal phase matrices. 
For simplicity, we have left out space-time arguments of the fields.
It is straightforward to check that invariance under this
transformation, using the Hamiltonians (\ref{HD}) and (\ref{HDmass})
and assuming non-vanishing neutrino masses, implies
\be\label{CPMUL}
\mbox{CP invariance} \quad \Rightarrow \quad
e^{i\alpha_R} \lambda^* e^ {-i\alpha_L} = \lambda, \quad
e^{i\alpha_R} M^* e^ {-i\alpha_L} = M
\quad \mbox{or} \quad
e^{i\alpha_L} U_L^* e^ {-i\alpha_R} = U_L \,.
\ee
Consequently, we can define phase-rotated quantities
\be\label{phaserot}
\nu_L = e^{i\alpha_L/2} \nu'_L, \; \nu_R = e^{i\alpha_R/2} \nu'_R
\quad \Rightarrow \quad 
\lambda' = e^{-i\alpha_R/2} \lambda e^{i\alpha_L/2}, \;
U_L' = e^{-i\alpha_L/2} U_L e^{i\alpha_R/2}
\ee
such that
\be
{U'_L}^* = U'_L \quad \mbox{and} \quad {\lambda'}^* = \lambda' \,.
\ee
As expected, CP invariance ensues the existence of phase-transformed
fields such that the mixing matrix $U_L'$ and the MM/EDM matrix 
$\lambda'$ are both real. 
Decomposing $\lambda'$ into MM and EDM matrices we obtain thus
\be\label{CPprop}
\mu' = \frac{1}{2} ( \lambda' + {\lambda'}^T ) \; 
\mbox{real, symmetric}, \quad
d' = \frac{i}{2} ( \lambda' - {\lambda'}^T ) \;
\mbox{imaginary, antisymmetric.}
\ee

If we go from the primed basis into the neutrino mass basis, we have 
$\tilde \lambda = \lambda' U'_L$ as MM/EDM matrix, which is again real.
Thus we have a decomposition of $\tilde \lambda$
into MM and EDM matrices with properties analogous to (\ref{CPprop}),
though the MMs and EDMs are related in a complicated way via
Eq.(\ref{newD}) with $S_L = U'_L$ and $S_R = 1$.

Let us now specialize this discussion to the physical situation of
neglecting neutrino masses in $\nu\, e^-$ scattering of solar neutrinos, 
such that neutrino masses enter only via the terms $M^\dagger M $ and
$M M^\dagger$ in the evolution equation (\ref{evol}).
Eq.(\ref{CPMUL}) leads to the following condition for CP invariance:
\be
M^\dagger M = e^{i\alpha_L} M^T M^* e^{-i\alpha_L}
\quad \mbox{and} \quad
M M^\dagger = e^{i\alpha_R} M^* M^T e^{-i\alpha_R} \,.
\ee
With the form of $M$ given in Eq.(\ref{HDmass}), the second relation is
trivially fulfilled and the first relation translates into
\be
U_L \hat{m}^2 U_L^\dagger =
e^{i\alpha_L} U_L^* \hat{m}^2 U_L^T e^{-i\alpha_L} \,.
\ee
The CP phases (\ref{CP}) of the right-handed fields do not occur in this
condition. Hence, a phase transformation of $\nu_R$ like in
Eq.(\ref{phasetrafoD}) can be used to remove $n$ phases from $\lambda$,
which introduces a change $\alpha_R \to \alpha_R - 2\sigma_R$ in the
CP transformation. Therefore, if, after performing the transformation
(\ref{phaserot}) on $\nu_L$, the mixing matrix has the form
$U'_L = V e^{i\hat \rho}$ with $V$ real and 
the MM/EDM matrix has the form $e^{i\hat\beta}\lambda'$
with $\lambda'$ real, where $\hat \rho$ and $\hat\beta$ are arbitrary
diagonal phase matrices, the Lagrangian is not invariant under CP in general.
However, the phases $\hat \rho$ and $\hat\beta$ do not lead to physical 
consequences in elastic $\nu\, e^-$ scattering of solar neutrinos.

Coming to CP invariance in the case of Majorana neutrinos, we use the
same CP transformation (\ref{CP}), except that the line with $\nu_R$
has to be dropped. The invariance of the mass term (\ref{HMmass}) requires
\be\label{CPmass}
e^{i\alpha_L} M e^{i\alpha_L} = -M^* \,.
\ee
Assuming now for simplicity not only non-zero but also non-degenerate neutrino
masses, one can show with Eqs.(\ref{HMmass}) and (\ref{CPmass}) that
the following conditions hold for CP invariance: 
\be\label{CPinvM}
U_L^* = ie^{-\alpha_L} U_L \varepsilon \quad \mbox{and}  \quad
e^{i\alpha_L} \lambda e^{i\alpha_L} = -\lambda^* \,,
\ee
where $\varepsilon$ is a diagonal sign matrix, which is not determined
by our manipulations. Making the phase redefinition (\ref{phaserot})
of the left-handed neutrino field, Eq.(\ref{CPinvM}) leads to 
\be\label{ULR}
U_L' = e^{-i\alpha_L/2} U_L = R\, e^{-i\varepsilon \pi/4}
\ee
for the mixing matrix, where $R$ is a real orthogonal matrix, and to
\be\label{l'}
\lambda' = e^{i\alpha_L/2} \lambda e^{i\alpha_L/2} \quad
\Rightarrow \quad {\lambda'}^* = -\lambda'
\ee
for the MM/EDM matrix. Consequently, with Eq.(\ref{l'}) we obtain
\be
\mu' = \lambda' \; \mbox{imaginary, antisymmetric}, \quad d' = 0 \,,
\ee
independent of the sign matrix $\varepsilon$.

Let us now relate the CP transformation in the flavour basis with
that in the mass basis. The mass eigenfields obtained by 
$\nu_L = U_L \tilde \nu_L$ have the matrix \cite{EGN}
\be
U_L^\dagger e^{i\alpha_L} U_L^* = i\varepsilon
\ee
instead of $e^{i\alpha_L}$ (\ref{CP}), where $U_L$ is given in
Eq.(\ref{ULR}). Hence, in the mass basis the 
CP transformation is given by \cite{wolf}
\be
\tilde{\nu}_L \to -i\varepsilon C \tilde{\nu}_L^* \,.
\ee
These CP signs $\varepsilon_j$ -- the CP parities -- then enter also
in the MM/EDM matrix in the mass basis given by
\be\label{ltilde}
\tilde \lambda = 
e^{-i\varepsilon \pi/4} R^T \lambda' R e^{-i\varepsilon \pi/4} =
\left( (R^T \lambda' R)_{jk} 
e^{-i(\varepsilon_j + \varepsilon_k) \pi/4} \right) \,.
\ee
If $\varepsilon_j = \varepsilon_k$, the phase factor in expression on 
the right-hand side of Eq.(\ref{ltilde})
is $\pm i$, whereas for $\varepsilon_j = -\varepsilon_k$ it is 1. In
the first case of equal CP parities, $\tilde \lambda_{jk}$ represents
a transition EDM, whereas for opposite CP parities this quantity is a
transition MM \cite{wolf,PSSV99}.
Therefore, for Majorana neutrinos and CP invariance in the primed flavour
basis one has $d'=0$, whereas in the mass basis either $\tilde \mu_{jk}$ or 
$\tilde d_{jk}$ is zero (or both are zero). This is 
completely different from the Dirac case where in both bases the same
properties (\ref{CPprop}) hold.

Let us now come to our physical approximation for elastic neutrino --
electron scattering with Majorana neutrinos. Eq.(\ref{CPmass}) implies for
the relevant term
\be\label{CPULMaj}
M^\dagger M = e^{i\alpha_L} M^T M^* e^{-i\alpha_L}
\quad \mbox{or} \quad
U_L \hat{m}^2 U_L^\dagger = e^{i\alpha_L} U_L^*\hat{m}^2 U_L^T
e^{-i\alpha_L} \,.
\ee
Obviously, the phase matrix $e^{-i\varepsilon\pi/4}$ 
drops out. Thus, in the case of CP invariance in our physical
scenario, the CP parities are irrelevant. Moreover, after performing
the phase transformation  
Eq.(\ref{phaserot}) on $\nu_L$, Eq.(\ref{CPULMaj}) is fulfilled for
$U'_L = R \,e^{i\hat\beta}$ with an arbitrary diagonal phase matrix
$\hat\beta$. Furthermore, there is the freedom to redefine $\nu_L$ with a
common phase which can be used to remove one phase from the MM/EDM
matrix. Therefore, if the mixing matrix has the form 
$R \,e^{i\hat\beta}$ and the MM/EDM matrix has the form 
$e^{i\gamma} \lambda'$ with ${\lambda'}^* = -\lambda'$, CP is violated
at the level of the Lagrangian in general. 
However, neither the phases $\hat\beta$ and
$\gamma$ nor the CP parities $\varepsilon$ lead to any physical 
consequences in our scenario.

\section{Decoherence effects as a consequence of neutrino energy averaging} 
\label{decoherence}

\subsection{Decoherence effects in the solar neutrino state}

In this section we consider the effect of neutrino oscillations and 
averaging over the neutrino energy in order to assess effective 
coherence or incoherence of the solar neutrino state arriving at the earth. 
We use the arguments presented, e.g., in Ref.\cite{DLS}.

The neutrino state undergoes only vacuum oscillations between
the sun and the earth. Therefore, denoting the values of 
$\varphi_\mp$ (\ref{evol}) at the edge of the sun by
$b_\mp$, we can write $a_\mp$ as
\be\label{ab}
a_- = U_L \exp \left( -i \hat{m}^2 L/2\omega \right) U_L^\dagger\, b_-
\,, \quad
a_+ = U_R \exp \left( -i \hat{m}^2 L/2\omega \right) U_R^\dagger\, b_+
\,,
\ee
respectively.
Here $L\approx 1.5 \times 10^{11}$ m is the distance between the sun and the
earth. Now the crucial point is that, according to the quadratic appearance
of $a_\mp$ in the cross sections (\ref{csWD}), (\ref{csWM}),
(\ref{csEM}), (\ref{csintD2}) and (\ref{csintM2}), the
following phase factors are important: 
\be\label{phase}
e^{\pm i\varphi_{jk}} \quad \mbox{with} \quad
\varphi_{jk} = 2\pi \frac{L}{\ell_{jk}} = 
\frac{\Delta m^2_{jk} L}{2\omega} \,,
\ee
where $\Delta m^2_{jk} = m^2_j - m^2_k > 0$ and
$\ell_{jk} = 4\pi \omega / \Delta m^2_{jk}$ is an oscillation length.
The phases (\ref{phase}) vary with energy as
\be
\delta \varphi_{jk} = \frac{\Delta m^2_{jk} L}{2\omega}
\frac{\delta\omega}{\omega}  =
2\pi \frac{L}{\ell_{jk}} \frac{\delta\omega}{\omega} \,.
\ee
Hence, integration over energy intervals such
$\delta\omega \gg  \omega \,\ell_{jk} / L$ $\forall j,k$
leads to an averaging of the oscillations, which can formally be
expressed as
\be\label{av}
\left\langle e^{\pm i\varphi_{jk}} \right\rangle = \delta_{jk} \,,
\ee
where $\delta_{jk}$ is the Kronecker delta.

Numerically, we have
\be\label{cohnum}
\frac{\ell_{jk}}{L} \approx 2.5 \: \frac{\omega (\mbox{MeV})}
{\Delta m^2_{jk} (\mbox{eV}^2) \: L (\mbox{m}) }
\approx 1.7 \times 10^{-11} \: \frac{\omega (\mbox{MeV})}
{\Delta m^2_{jk} (\mbox{eV}^2)}
\approx
\frac{4.5 \times 10^{-12} }{\Delta m^2_{jk} (\mbox{eV}^2)} \,,
\ee
where in the last step we have used $\omega \approx 0.27$ MeV,
the average energy of the
$pp$-neutrinos, which are most suitable for measuring the azimuthal
asymmetry in $\nu e^-$ scattering \cite{PSSV98}. If we consider, for example,
$\Delta m^2 \sim 10^{-8}$ eV$^2$ allowed by the RSFP scenario
\cite{LM,Akhmedov,minakata,BPRS,GN,DT,PA}, 
we find $\ell /L \sim 5 \times 10^{-4}$, where
$\ell$ is the oscillation length corresponding to $\Delta m^2$. Therefore,
to avoid the averaging (\ref{av}) associated with the vacuum oscillations,
one would have to measure the neutrino energy with an accuracy better than
$\delta \omega /\omega \sim 10^{-4}$, which seems rather impossible.

Actually, even if we concentrate on the solar $^7$Be line with $\omega
= 862.27$ keV \cite{bahcall}, the natural line broadening by the high
temperatures in the center of the sun with $\delta \omega = 1.63$ keV
\cite{bahcall} is sufficient to cause considerable averaging. In this
case we obtain $(L/\ell)(\delta \omega/\omega) \approx \Delta m^2/7.8
\times 10^{-9}\, \mbox{eV}^2$ and, therefore, a decoherence effect for 
$\Delta m^2 \gtrsim 10^{-8}$ eV$^2$ \cite{fogli}.

The energy averaging of the vacuum
oscillations is equivalent to consider the neutrino
state arriving at the earth as an incoherent mixture of mass eigenstates.
In the case of total incoherence the density matrix is a diagonal matrix
in the mass basis: $\rho_{rs} = \mbox{diag}(a_r^j a_s^{j\ast})$ with
$r,s = \pm$, where $j$ numbers the neutrino mass eigenstates.

\subsection{The energy-averaged cross sections}

In this and the next subsection we assume that $\omega$ represents an average
neutrino energy or the center value of an energy interval of length 
$\delta \omega$ over which the averaging takes place. 
Furthermore, we assume that $\delta \omega \ll \omega$ holds 
and that the averaging condition $\delta \varphi_{jk} \gg 2\pi$ 
holds for all neutrino masses $m_j \neq m_k$. 
Performing the averaging procedure
(\ref{av}) in the weak, electromagnetic and interference cross sections, it
turns out that the averaged cross sections are written in a simpler way
by using the coefficients 
\be\label{b}
\tilde{b}_- = U_L^\dagger b_- \,, \quad 
\tilde{b}_+ = U_R^\dagger b_+ \,,
\ee
and the matrix
\be\label{l}
\tilde{\lambda} = U_R^\dagger \lambda U_L \,,
\ee
which represent the flavour coefficients
$b_\mp$ (\ref{ab}) and the matrix $\lambda$ (\ref{lambda}), respectively, 
transformed into in the mass basis. Eqs.(\ref{b}) and (\ref{l}) refer to the
Dirac case, but in the Majorana case one only has to replace
$U_R$ by $U_L^\ast$ (see Eq.(\ref{diff})).
In the following we use the notation $\tilde{b}_\mp^T = (b_\mp^j)$, i.e., 
we label flavour indices with $\alpha$ and mass indices
with $j$. Using Eqs.(\ref{b}) and (\ref{l}), after neutrino energy
averaging or for effective total coherence loss between neutrino mass
eigenstates, the cross sections (\ref{csWM}), (\ref{csEM}) and
(\ref{csintM2}) for Majorana neutrinos take the shape
\begin{eqnarray}
\left\langle \frac{d^2 \sigma^M_\mathrm{w}}{dTd\phi} \right\rangle
&=& \sum_{\alpha,j} |U_{L\, \alpha j}|^2
\left( |b_-^j|^2 \frac{d\sigma(\nu_\alpha e^-)}{dTd\phi} +
|b_+^j|^2 \frac{d\sigma(\bar{\nu}_\alpha e^-)}{dTd\phi}
\right) \,, \label{csWmassin}\\
\left\langle \frac{d^2 \sigma_\mathrm{em}}{dT d\phi} \right\rangle
&=& \frac{\alpha^2}{2m_e^2 \mu_B^2}
\left( \frac{1}{T} - \frac{1}{\omega} \right) \sum_j
\left(
|b_-^j|^2 (\tilde{\lambda}^\dagger \tilde{\lambda})_{jj} +
|b_+^j|^2 (\tilde{\lambda} \tilde{\lambda}^\dagger)_{jj}
\right) \,, \label{csEMmassin} \\
\left\langle \frac{d^2 \sigma^M_\mathrm{int}}{dT d\phi} \right\rangle
&=& F \, \mbox{Re}\left[ \sum_j b_+^{j\ast} b_-^j
\left( \tilde{\lambda} U_L^\dagger (g - \bar{g}) U_L \right)_{jj}
(p_x' - ip_y')\right] \,, \label{csintmassin}
\end{eqnarray}
respectively.
The corresponding expressions for Dirac neutrinos are obtained from
Eqs.(\ref{csWmassin}) and (\ref{csintmassin}) by dropping the 
$\sigma(\bar{\nu}_\alpha e^-)$ and $\bar g$ terms, respectively.

It is interesting to note that for total incoherence of the neutrino mass
eigenstates only the
axial part of the weak interaction contributes to the interference
cross section for Majorana neutrinos. Inserting Eq.(\ref{defg}) into 
the cross section Eq.(\ref{csintmassin}), we obtain
\be\label{csintmassin2}
\left\langle \frac{d^2 \sigma^M_\mathrm{int}}{dT d\phi} \right\rangle
=2 F\,\frac{T}{\omega} \, \mbox{Re}\left[ \sum_{\alpha,j,k}
b_+^{j\ast} b_-^j \, \tilde{\lambda}_{jk} \, 
U^\ast_{L\, \alpha k} U_{L\, \alpha j}
\, g^\alpha_A \, (p_x' - ip_y')\right] \,.
\ee
The dependence on the electron recoil energy of this expression is very
different from the corresponding term (\ref{csintM2})
in the case of full coherence and the Dirac terms with and without coherence
(see Eqs.(\ref{csintD2}) and (\ref{csintmassin}) without the 
$\bar g$ term), because the recoil energy $T$ drops out of the product $FT$.

\subsection{Decoherence in the 2-Majorana neutrino case}

Now we consider in detail the effect of decoherence for the
two flavours $e$ and $x=\mu,\tau,s$ of Majorana neutrinos. For this purpose we
will refer to the discussion in Section \ref{2M}. We have proved in this
section that all phases of the problem are unphysical. Therefore, we use the
mixing matrix (compare with Eq.(\ref{UL}))
\be\label{V}
V = \left( \begin{array}{rr} c & s \\ -s & c \end{array} \right) \,,
\ee
where $c \equiv \cos\theta$, $s \equiv \sin\theta$ and the quantity 
$| \Lambda |$ for the transition MM/EDM (see Eqs.(\ref{2Mlambda}) and
(\ref{Lambda})). With the averaged Majorana interference cross section
(\ref{csintmassin2}) and
\be\label{lcs}
\tilde \lambda = | \Lambda | V^T \epsilon V = | \Lambda | \epsilon \,,
\ee
we arrive at the final result
\be\label{csint2maj}
\left\langle \frac{d^2 \sigma^M_\mathrm{int}}{dT d\phi} \right\rangle =
F| \Lambda | \sin 2\theta\,  \frac{T}{\omega}\, (g_A^e - g_A^x)\,
\mbox{Re} \left[ \left( 
(b_{\hphantom{\prime\hskip+0.5pt}+}^{\prime\, 1})^* 
 b_{\hphantom{\prime\hskip+0.5pt}-}^{\prime\, 1} -
(b_{\hphantom{\prime\hskip+0.5pt}+}^{\prime\, 2})^* 
 b_{\hphantom{\prime\hskip+0.5pt}-}^{\prime\, 2}
\right) (p_x' - ip_y') \right] \,.
\ee
Note that the vectors 
$\tilde{b}'_\mp = ( b_{\hphantom{\prime\hskip+0.5pt}\mp}^{\prime\, j})$ 
represent the neutrino state at the edge of the sun: by Eq.(\ref{b})
they are related to the flavour vectors $b'_\mp$ which
are obtained by evolution with the Hamiltonian
$H'_\mathrm{eff}$ (\ref{Heff'}) -- 
analogous to $a'_\mp$ (\ref{a'}) -- and are
independent of any phases initially in $U_L$ and $\lambda$.

The expression (\ref{csint2maj}) is proportional to the mixing angle 
$\sin 2\theta$. This shows that the question if the solar neutrino state on
earth is to be considered as a coherent or effectively incoherent admixture of
mass eigenstates has a strong effect on the interference cross section, whereas
this question has no bearing on the electromagnetic cross section in the
2-Majorana case. Large values for $\sin 2\theta$ are disfavoured in the RSFP
scenario \cite{ALP,BPRS} and by the
non-observation of electron antineutrinos in Super-Kamiokande 
\cite{barbieri,BPRS,SPV,TL} and  hence the asymmetry is suppressed.
These arguments suggest that a significant asymmetry measured
in an experiment is unlikely to result from a 2-Majorana neutrino
scenario, except for very small mass-squared differences
($\Delta m^2 < 10^{-11}$, see Eq.(\ref{cohnum})). Of course, it could
result from Dirac diagonal moments. In this case
the states of negative and positive helicity belong to the same
mass eigenvalue and no averaging due to oscillations is possible.

\section{Conclusions}
\label{conclusions}

In this paper we have considered elastic neutrino -- electron
scattering of solar neutrinos, taking into account the possibility that
neutrinos have MMs and EDMs. We have presented the most general
cross section for an initial neutrino state, which can be an arbitrary
superposition of different neutrino types -- including sterile
neutrinos -- with arbitrary helicities. Consistency requires that the
neutrino superposition which undergoes the elastic neutrino --
electron scattering is considered as the result of an evolution of the
initial electron neutrino state with negative helicity generated in the core
of the sun. The neutrino MMs and EDMs enter into this evolution
equation (\ref{evol}) as well as into the cross section. Only by
taking this 
twofold effect of the MM/EDM matrix (\ref{lambda}) into account, the
final results for the pure electromagnetic cross section (\ref{csEM}) 
and the weak-electromagnetic interference cross section, i.e.,
(\ref{csintD2}) for Dirac and (\ref{csintM2}) for Majorana neutrinos,
are invariant under phase transformations (\ref{phasetrafoD}) and
(\ref{phasetrafoM}) of the neutrino fields. 

In this context we have to mention our approximation:
We have neglected neutrino masses in the cross section, but in the
evolution equation (\ref{evol}) we have taken
into account the usual quadratic dependence of the effective Hamiltonian
on the neutrino masses, as required by the effect of
background matter \cite{matter}. We have formulated the cross section
and the evolution equation in the flavour basis. However, we want to
stress that we are not obliged to stick to this basis. Since we
neglect neutrino masses in the cross section, we could choose rotated
neutrino fields according to Eq.(\ref{unitary}). The final result for
the cross section would not depend on the transformation matrices 
$S_L (S_R)$ (see Subsection \ref{invariance}). 
Thus, with our physically motivated approximation there
is no preferred basis of neutrino fields.

Of particular importance is the
weak-electromagnetic interference cross section: if it were non-zero,
it would indicate that the solar neutrinos have acquired some amount of
transverse polarization due to MMs and EDMs and a magnetic field in
the solar interior. Such an interference cross section would
show up in an azimuthal asymmetry of the momentum distribution of the recoil
electron, in the plane orthogonal to the direction of the
incoming neutrino \cite{BF}. 

In our physical scenario we have shown that
in the 1-Dirac neutrino case the azimuthal asymmetry does not allow to
distinguish between MM and EDM but is a function of $\sqrt{\mu^2+d^2}$
(\ref{aa}) as is the pure electromagnetic cross section. 
In the 2-Majorana neutrino case the same holds for the transition
moments (see Eq.(\ref{int})) and, in addition, none of the phases in the
neutrino mixing matrix $U_L$ is physical either. We have also made a
general counting of the physical, independent phases in our framework for $n$
neutrino flavours or types in the Dirac and Majorana cases. We have
pointed out that, by phase transformations on the neutrino fields,
phases can be shifted from $U_L$ to $\lambda$ and vice versa such that
in order to obtain phase convention-independent quantities one has to
combine elements from both matrices according to the phase
transformations (\ref{phasetrafoD}) and (\ref{phasetrafoM}).
We want to stress that the entity which transforms under basis
transformations in correspondence with the neutrino fields is the
matrix $\lambda = \mu - id$, but not the separate MM and EDM matrices
$\mu$ and $d$, respectively. Furthermore, in the
flavour basis we are working with, for Majorana neutrinos 
the so-called Majorana phases drop out trivially, because in our approximation
the neutrino mass matrix appears only in the evolution equation (\ref{evol}) as
$M^\dagger M$ and $M M^\dagger$ (see Eqs.(\ref{masses}) and (\ref{UL})). 
The same holds, in the case of CP invariance, for the CP parities 
of the neutrino mass eigenfields. 

Finally, we have shown that averaging over small neutrino energy
intervals, which is inevitable through realistic neutrino detection, has a
drastic effect on the weak-electromagnetic interference cross section.
This effect comes about because of neutrino oscillations in
vacuum between the sun and the earth \cite{DLS} and is operative at
least for neutrino mass-squared differences larger than about 
$10^{-9} \div 10^{-10}$ eV$^2$, having in mind solar neutrino 
energies below 1 MeV. In the 2-Majorana neutrino case the
averaged interference cross section is then proportional to $\sin
2\theta$, where $\theta$ is the mixing angle in $U_L$. Thus, in a
2-Majorana RSFP scenario without mixing, the averaged interference
cross section is zero. However, mixing in the general 2-Majorana 
RSFP scenario tends to be suppressed anyway according to the
non-observation of solar $\bar\nu_e$'s in Super-Kamiokande. 
An observation of a significant azimuthal asymmetry could be an indication
of very small mass-squared differences or of Dirac diagonal moments.

\acknowledgments
We thank V.B. Semikoz and J.W.F. Valle for useful discussions.

\end{document}